\documentclass[a4paper]{article}
\usepackage[margin=2cm]{geometry}
\usepackage{graphicx}

\begin{document}

\title{An adaptive music generation architecture for games\\
based on the deep learning Transformer model}

\author{Gustavo Amaral Costa dos Santos$^1$
	\and Augusto Baffa$^1$
	\and Jean-Pierre Briot$^{2, 1}$\\
	\and Bruno Feijo$^1$
	\and Antonio Luz Furtado$^1$}

\date{$^1$ Dept. of Informatics, PUC-Rio, Rio de Janeiro, Brazil\\
	$^2$ Sorbonne Universit\'e, CNRS, LIP6, F-75005 Paris, France\\[.2cm]
	{\tt\small gustavoacs99@gmail.com}, {\tt\small  abaffa@inf.puc-rio.br}, {\tt\small  Jean-Pierre.Briot@lip6.fr},\\
	{\tt\small  bfeijo@inf.puc-rio.br}, {\tt\small  furtado@inf.puc-rio.br}}

\maketitle

{\bf Abstract:}
This paper presents an architecture for generating music
for video games based on the Transformer deep learning
%technique.
model.
Our motivation is to be able to customize the generation according to the taste of the player,
who can select a
%set
corpus
of training examples,
corresponding to his preferred musical style.
%of music.
The system generates various musical layers,
%to apply
following the standard layering strategy currently used by composers designing video game music.
To adapt the music generated to the game play and to the player(s) situation,
%he music is adaptive to the psychological context of the player,
we are using an arousal-valence model of emotions, in order to control the selection of musical layers.
We discuss current limitations and prospects for the future,
such as collaborative and interactive control of the musical components.


\section{Introduction}
\label{section:introduction}

Music is
%very important
essential
in video games. It provides an embedding context for the players and complements the scenario \cite{sanders:perception:music:video:games:bcs:2010}.
%\cite{lima:music:director:sbgames:2020}.
%It
Music
can also offer some ways of controlling the player \cite{jacopin:music:tempo:tetris:iui:2021}.
Meanwhile, a recent observation is that many players replace
%the music of a game
a game's music
by listening to some
%music
musical piece
of their choice \cite{ramalho:game:music:earphones:2021}.
We postulate that this is because
of
the absence of enough personalization of the music
%offered by the game.
the game offers.
Therefore, we started
%to investigate
investigating
the possibility of generating personalized music
based on player preferences (music style, as defined by a corpus of music samples).

Deep learning techniques
are effective in learning a
music
style from a corpus and generating conformant samples
%{\footnote{An example, among many,
%	is OpenAI MuseNet \cite{payne:musenet:blog:2019}.}.
\cite{dlt4mg:springer:2019}.
%Still remain the
Various issues still remain
%about
on
how to customize,
control
and orchestrate
the generation
of music
in
%the
function of the situation (game and player).
%In the following, we will introduce the design, implementation
%and preliminary experiments with a prototype architecture for
%generating personalized and adaptive music for games.
%However,
Because such various open questions,
%at a time of so few works in game music generation
and the fact that generative music for games is still a recent
%and open
domain
with few prototypes
(as surveyed in \cite{plut:music:games:ec:2020} and Section~\ref{section:adaptative:generative}),
%with
%and so
%many
we believe in the importance of searching for simple models
that explore
%more
fundamental aspects of
%generating music for games.
%video
music generation for games.
More specifically, it becomes essential to look for models that, more straightforwardly,
represent the modes of emotion and levels of emotional intensity involved in video game music generation.
Also, we must look for deep learning techniques that are especially adequate to the musical narrative.
And, perhaps even more importantly, we should seek a model to support instrumental layers used in video game composition
(such as the practice of ``striping'',
which is to record orchestral sections separately for future mixing according to the whims of the composer).
As
    film and game music composer
    E\'{\i}mear Noone explains:
    ``we might record the strings separately, for example,
    but we'll compose in a way that the strings on their own
    provide a functioning piece of music.
    Then, if our character triggers something in the world,
    perhaps a battle,
    we can land the wood winds or brass on top of that
    to increase the intensity.
    Each part must be self-contained yet work with others --
    you need to be able to kick in the brass,
    kick in the percussion, whenever it's triggered by gameplay.'' \cite{stuart:video:game:music:theguardian:2019}.
    See also, e.g.,
    \cite{layering:hyperbits:2022}
    for a general introduction to layering.

In this context, instead of looking for alternatives or improvements
in the few existing complete models for
%adaptive music composition
game music generation
(such as the excellent work by Hutchings and McCormack
	\cite{hutchings:amc:trans:2020},
	to be analyzed in Section~\ref{section:architecture:ams}),
we decided to explore more straightforward and flexible models for generating and adapting music, based on layering \cite{layering:hyperbits:2022}.
Also, we believe that our proposed model can facilitate the control and orchestration of music for video games in a collaborative environment.
%In the following, we will introduce the design, implementation
%and preliminary experiments with a prototype architecture for
%generating personalized and adaptive music for games.

With the above mentioned principles in mind, after several experiments,
we opted for the Transformer architecture
\cite{vaswani:attention:transformer:arxiv:2017},
because it better captures the long-term
%coherence
%and
structure
of music \cite{huang:music:transformer:arxiv:2018}.
%Our following insight proposed a two-dimensional space of emotions
%to simplify the metric space and help the practice of layering music.
%For that, we adopt

In order to model the psychological state of the player as respect to the gameplay context,
we decided to select
the relatively standard arousal/valence model of emotions
\cite{russell:circumplex:model:1980},
and to map it to the control (adaptation) of the generation of music.
%In this model,
%we
We
associate arousal (i.e., intensity) with the number of active layers
(e.g., the system can add a layer with woodwinds or brass to increase the intensity level
if a battle starts in the game's world).
And the valence corresponds to the emotional modes of the generated music.
We also discuss, in this article, future extensions that this simplified approach makes easier to implement.
In particular, we want to move towards collaborative and interactive control of the music components generated by the Transformer-based architecture.

The following sections introduce the design, implementation
and preliminary experiments with a prototype architecture
for generating personalized and adaptive music for games aligned with the above mentioned principles.

\section{Background and Related Work}
\label{section:related:work}

\subsection{Adaptive versus Generative}
\label{section:adaptative:generative}

In \cite{plut:music:games:ec:2020}, Plut and Pasquier present a survey about various approaches and challenges for the generation of music for video games.
They consider two primary techniques:

\begin{itemize}

\item {\em adaptive music} (also named {\em interactive music}),
where music is organized in order to be able to react to a game's state \cite{collins:interactive:audio:games:2011}.
Some musical features (e.g., adding or removing instrumental layers
%\footnote{An example of adapting instrumental layers is a recording method known as striping: ``we might record the strings separately, for example, but we'll compose in a way that the strings on their own provide a functioning piece of music.
%Then, if our character triggers something in the world, perhaps a battle, we can land the wood winds or brass on top of that to increase the intensity.
%Each part must be self-contained yet work with others ? you need to be able to kick in the brass, kick in the percussion,
%whenever it's triggered by gameplay'' as E\'{\i}mear Noone explains, cited in \cite{stuart:video:game:music:theguardian:2019}.
%Also, see, e.g.,
%\cite{layering:hyperbits:2022}
%for a general introduction to layers.},
(such as for striping),
changing the tempo, adding or removing processing, changing the pitch content\ldots)
are linked to game play variables.

An example of adaptive music is the ``Luftrausers'' game \cite{luftrausers:2014},
where the composed music has been split into 3 groupings of instruments, each of which has 5 different arrangements,
which a player may select for his avatar (see more details in \cite[Section 1.2]{plut:music:games:ec:2020}).

\item {\em generative music},
where music is not preexisting and dynamically adapted, as for adaptive music, but is generated on the fly.
It is created in some systemic way by the computer and is sometimes called procedural music or algorithmic music \cite{nierhaus:algorithmic:composition:book:2009}.
The musical content is generated from some model.

The model can be specified by hand.
This was for instance the case for the first piece of music composed in 1957 by a computer
(the ``ILLIAC I'' computer at the University of Illinois at Urbana-Champaign (UIUC) in the United States),
    and therefore named ``the Illiac Suite'' \cite{lejaren:illiac:book:1959}.
    The human ``meta-composers'' were
%   Lejaren A.
    Hiller and
%   Leonard M.
    Isaacson, both musicians and scientists.
    It was an early example of algorithmic composition,
    making use of stochastic models (Markov chains) for generation 
    as well as rules to filter generated material according to desired properties.
The limits are that specifying the model is difficult and error prone.
The progress of machine learning techniques
made it possible to learn models from examples (in other words, specify a model by extension rather than by intention).
All but one of the generative music systems surveyed in \cite{plut:music:games:ec:2020} are using Markov chains models.
%(e.g., as the system presented in \cite{engels:aaai:2015}).
Markov models are indeed simpler than deep learning/neural networks models,
but they face the risk of plagiarism,
by recopying too long sequences from the corpus.
Some interesting solution is to consider a variable order Markov model and to constrain the generation (through min order and max order constraints) \cite{papadopoulos:maxorder:universality:book:2016}.
%\cite[Section 1.2.3]{dlt4mg:springer:2019}.
The only surveyed game music generation system based on artificial neural networks is Adaptive Music System (AMS)
by Hutchings and McCormack \cite{hutchings:amc:trans:2020}
and it will be summarized in next section (Section~\ref{section:architecture:ams}).

\end{itemize}

Generative music is more general and adaptive than pre-composed composed adaptive music,
but is also more difficult to control and more computing demanding.
%, specially in order to generate music on demand.
As, noted by \cite{plut:music:games:ec:2020}: ``Another reason that generative music may not have received
widespread attention in the games industry is that it is often unpredictable
and can be difficult to control.
The audio director of ``No Man's Sky'' game, Paul Weir, notes that generative music was used in the
game with an acknowledgment that it could produce ``worse'' music than composed music.'' \cite{weir:no:man:sky:2017}.
Actually, that distinction between adaptive and generative music is not that clear, as often systems classified as generative are not completely generative
and include adaptation components.
This is for instance the case of the Adaptive Music System.
Note that it is classified as generative in \cite{plut:music:games:ec:2020},
	although its very name claims it as adaptive!
	In fact we need systems to be both generative and adaptive.
%(AMS) \cite{hutchings:amc:trans:2020}.
We will now summarize it in next section (Section~\ref{section:architecture:ams}) in order to illustrate some issues and also for its own merits.

\subsection{Architecture of the Adaptive Music System}
\label{section:architecture:ams}

\begin{figure}[htbp]
\begin{center}
\includegraphics[width=15cm]{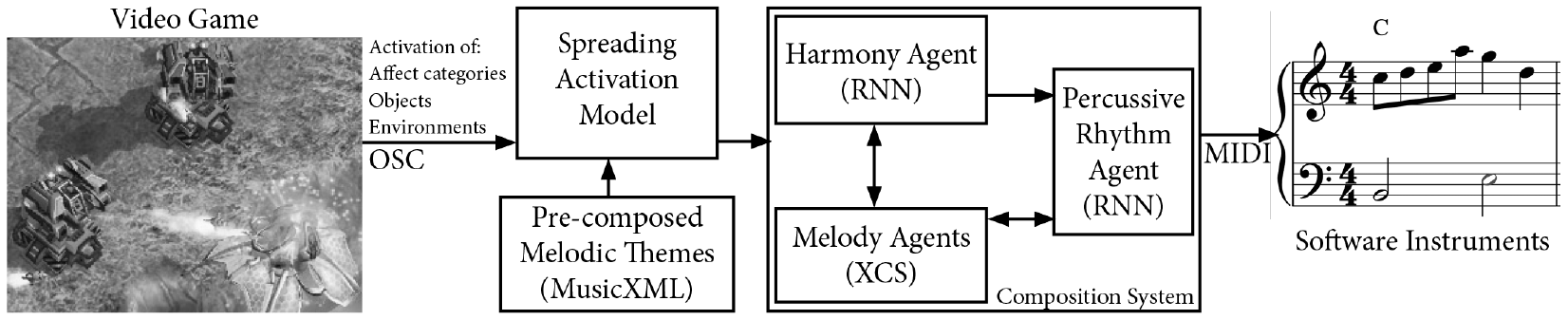}
\caption{AMC architecture, reproduced from \cite{hutchings:amc:trans:2020}}
\label{figure:amc:architecture}
\end{center}
\end{figure}

The architecture of (AMS)
\cite{hutchings:amc:trans:2020}
(illustrated in Fig.~\ref{figure:amc:architecture})
is multi-agent and multi-technique:

\begin{itemize}

\item the {\em harmony role agent}, which generates a chord progression, using an RNN (trained on a corpus of symbolic chord sequences, actually an extension of the harmony system from the same authors
\cite{hutchings:improvise:2017});

\item the {\em melody role agents} (one for each instrument), which instantiate characteristics (length, pitch, proportion of diatonic notes\ldots) of pre-existing melodies,
%\footnote{In some ordering strategy:
%    the first melody agent produces the melodic line with the highest pitch;
%    while the second melody agent produces the lowest pitch melodic line;
%    and subsequent melody agents gradually get higher above the lowest melody line
%    until the highest melody line is reached.
%    This strategy is loosely inspired by counterpoint harmonization composition
%    techniques.},
using an evolutionary rule system (XCS, for eXtended learning Classifier System
\cite{wilson:xcs:1995}) and adapting them to the harmony;
%\footnote{Actually, the strategy is more democratic and dynamic,
%    as harmony agent or melody agents can become leaders
%    depending on their own confidence,
%    based on the output of the softmax layer for the harmony agent prediction.};

\item the {\em rhythm role agent}, which uses another RNN model.

\end{itemize}

AMS considers a model of 6 emotions: sadness, happiness, threat, anger, tenderness and excitement\footnote{The 5 first ones are
    the most consistently used labels in describing music across multiple music listener studies \cite{juslin:music:emotions:2013}.
    Excitement has been added as an important aspect of emotion for scoring video games.},
whose selection is triggered by current game state (every 30ms, the list of messages received by the Open Sound Control (OSC) is used to update the activation values).
Emotions in turn will modulate the selection among
%pre-melodies
melodies (choosing the melodic theme assigned to currently highest activated concept or affect),
%-- see in next footnote information about the spreading activation model)
with the instantiation of their characteristics
%(using a predefined conversation table, shown in Fig.~\ref{figure:correlation:table}),
being managed by a spreading activation model.
It is
%   As explained in \cite{hutchings:amc:trans:2020}:
%   "This model is constructed as a graph of concept nodes that are connected by weighted edges representing the strength of the association between the concepts.
    a graph of concept nodes, connected by weighted edges representing the strength of the association between the concepts
    (and is inspired from
%    the spreading activation model of
    a semantic content organisation in cognitive science
%   from
    \cite{collins:activation:1975}).
%    When a person thinks of a concept, concepts connected to it are activated to a degree proportional to the weight of the connecting edge.
    Activation spreads as a function of the number of mediating edges and their weights.
%    and "Spreading activation models don't require logical structuring of concepts into classes or defining features,
%    making it possible to add content based on context rather than structure.
    As explained in
    \cite{hutchings:amc:trans:2020}:
    ``Spreading activation models don't require logical structuring of concepts into classes or defining features,
    making it possible to add content based on context rather than structure.
    For example, if a player builds a house out of blocks in Minecraft, it does not need to be identified as a house.
    Instead, its position in the graph could be inferred by time characters spend near it, activities carried out around it, or known objects stored inside it."

As we can see, AMS actually proposes a sophisticated (and clever) generation model
which includes both adaptive and generative aspects (for instance, harmony is generated and melodies are adapted).
AMS has been tested in two games: Zelda Mystery of Solarus (MoS) (actually an open-source clone version)
\cite{zelda}
and StarCraft II
\cite{starcraft:ii}.

The comparison of our proposed model with the much more complete and robust AMS architecture is twofold.
Firstly, we more straightforwardly represent the emotional intensity in music generation,
and secondly, we better support layering music.
The simplicity of our approach aims to facilitate future prospecting in collaborative and interactive environments. Furthermore, we use a deep learning architecture (Transformer)
better suited to capture long-term coherence in music.

%\begin{figure}[htbp]
%\begin{center}
%\includegraphics[width=8cm]{correlation-table.png}
%\caption{Correlation table between music melody properties and emotion category, reproduced from \cite{hutchings:amc:trans:2020}}
%\label{figure:correlation:table}
%\end{center}
%\end{figure}

\section{Adaptability versus Continuity and other Design Issues}
\label{section:adaptibility:continuity}

A pure generative approach is some kind of ideal, as it could in principle combine personalization (learnt styles) with real-time adaptation (to the game and players situation).
Note that
the issue of how to combine
    various context information, plot, evolution, player(s) situation, etc., including statistics, e.g., average reactivity of a player, into some decision about what is the objective
    (adapt to current game context, or the opposite, trigger a player to engage more) and how to accordingly adapt the music is still an open issue.
    It is likely that it should use some aggregation/interpretation rules, as well as multi-criteria decision strategy,
    within some front end module in charge of mapping events and models from the game up to the control parameters for music generation or/and adaptation.

Using symbolic-level music models as opposed to signal-level music models brings the advantages of higher level manipulation (at the composition level) and less computer resources
(although,
    recent waveform-level models such as WaveNet \cite{oord:wavenet:arxiv:2016}
    demonstrated the feasibility of real-time conditioned generation, used for instance for intelligent assistants such as Google Echo or Amazon Alexa).
An important and actually difficult issue remains the capacity to generate on the fly music content\footnote{Hard real-time may be unnecessary,
    as the music does not have to adapt immediately to events,
    as opposed to sound effects.}
and be able to adapt it, while maintaining some continuity
(as for a musician improvising in some jazz context, balancing between constructing and following some musical discourse and fitting into the dynamic context, in the first place, harmonic modulations).
Recent progress for control strategies for Markov chains and as well for deep learning show promising results.
Markov constraints have been proposed as an unifying framework for Markov-based generation while satisfying constraints \cite{pachet:markov:constraints:constraints:2011},
and has been applied to real-time improvisation \cite{pachet:virtuoso:2012} and to interactive composition \cite{papadopoulos:flow:composer:cp:2016}.
Challenges for introducing control are somehow harder for deep learning,
(as explained, e.g., in
%[Hidden Reference]),
\cite[Section~10]{nn4music:ncaa:2020}),
but progresses are made, using control strategies such as conditioning
(adding some additional input to the neural network in order to parameterize training and generation),
e.g., as in \cite{hadjeres:anticipation:rnn:arxiv:2017}.

\section{Current Proposal}
\label{section:current:proposal}

%Our current prototype,
%\cite{amaral:graduacao:2021},
Although simpler,
our current prototype
shares some similarity with AMS
(see Section~\ref{section:architecture:ams} and \cite{hutchings:amc:trans:2020}),
in that it uses both neural network-based generation and
%also
an emotion reference model.
%However, we are using the recent Transformer deep neural network architecture %\cite{vaswani:attention:transformer:arxiv:2017},
%which recently became popular for such applications as: translation, text generation (e.g., the Generative %Pre-trained Transformer 3 aka GPT-3 model), biological sequence analysis,
%and music generation \cite{huang:music:transformer:arxiv:2018}.
%by better handling long-term correlations.
In the following sub-sections, we will describe and motivate various aspects and components of the architecture and of the generation process,
namely:
the general design principles;
the curation and pre-processing of the training musical examples;
the way music generated is layered;
the emotion model chosen to map the game play into some control of the generated music;
the mapping discipline;
the complete architecture;
the implementation;
and the preliminary evaluation.

\subsection{Design Principles}

\begin{figure}[htbp]
\begin{center}
\includegraphics[width=8cm]{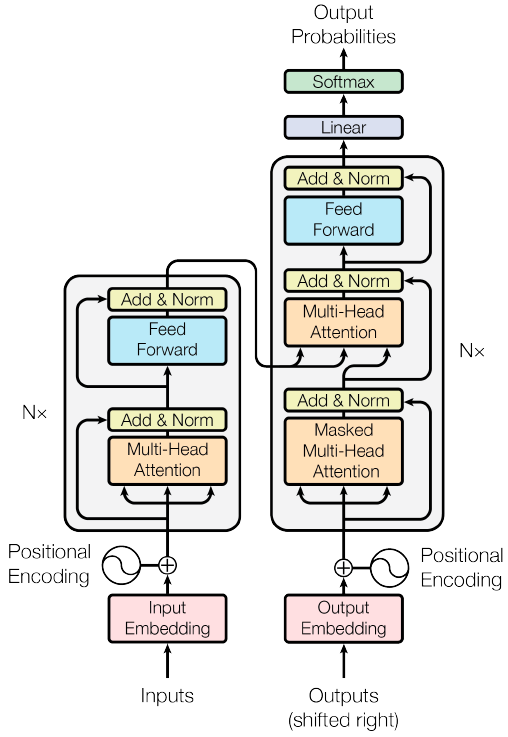}
\caption{Transformer architecture, reproduced from \cite{vaswani:attention:transformer:arxiv:2017}}
\label{figure:transformer}
\end{center}
\end{figure}

%The main result is the design and implementation of a first prototype
%(proof of concept) of a music generation architecture for games,
%based on the Transformer architecture \cite{vaswani:attention:transformer:arxiv:2017}.
After having at first experimented with a recurrent neural network architecture of type LSTM (part of Google's Magenta project library) \cite{magenta},
we selected the Transformer architecture
for its ability
to
%%deal with long term correlations,
%and therefore
enforce
%larger scale
consistency and structure,
by better handling long-term correlations.
Transformer
\cite{vaswani:attention:transformer:arxiv:2017}
is an important evolution
of a Sequence-to-Sequence architecture
(based on RNN Encoder-Decoder), where a variable length sequence
    is encoded into a fixed-length vector representation which serves as a
    pivot representation to be iteratively decoded to generate
    a corresponding sequence
    (see more details,
    e.g., in
%    \cite[Section~5.13.3]{dlt4mg:springer:2019}).
    \cite[Section~10.4]{goodfellow:deep:learning:book:2016}).
    Its main novelty is a self-attention mechanism
    (as a full alternative to more classical mechanisms
    such as recurrence or convolution),
%   which allows
    to focus on contributing elements of an input sequence.
    For more details on the architecture,
    illustrated in Fig.~\ref{figure:transformer},
    please see the original article
    \cite{vaswani:attention:transformer:arxiv:2017}
    or some pedagogical introduction
    \cite{phi:transformer:towards:data:science:2020}).
It
recently became popular for such applications as: translation, text generation (e.g., the Generative Pre-trained Transformer 3 aka GPT-3 model), biological sequence analysis
and music generation \cite{huang:music:transformer:arxiv:2018}.
%It is based on a Sequence-to-Sequence architecture
%\footnote{It is based on a RNN Encoder-Decoder architecture,
%   where two RNN architectures are nested within the encoder and decoder parts of
%   an autoencoder.
%    The idea is to encode a variable length sequence learnt by a recurrent network
%    into a fixed-length vector representation which serves as a
%    pivot representation to be iteratively decoded to generate
%    a corresponding sequence.
%    The motivation and application target has been translation
%    from one language to another,
%    resulting in sentences of possibly different lengths.}
%\cite{sutskever:sequence:2:sequence:nips:2014},

The proposal by Jeffries for ambient music generation based
on the Transformer
\cite{jeffries:musician:machine:blog:magenta:2020}
has also been a source of inspiration.
%\cite{whitaker:ambient:generator:2020}.

\subsection{Training Examples}

%\footnote{The main motivation
%    is the personal preference of the 1st author.},
%bearing in mind the generality of the model,
The user may select a corpus of music of its preference
(e.g., classical, jazz, techno, ambient\ldots,
choosing a more narrow -- e.g., of a specific composer or band -- or some wider corpus)
to be used as the set of training examples.
The music generated will be corresponding to this style thus defined by the user.
In the experiment described in this paper,
we have chosen a corpus of ambient music,
more precisely a Spotify playlist named ``Ambient songs for creativity and calm'',
curated by
%Dan
Jeffries, and containing approximately 20 hours and 165 titles
\cite{jeffries:ambient:playlist:2022}.
 
The compressed audio files (mp3) corresponding to the musical training examples
have been uncompressed into waveform (wav) files and then, by using a pitch detector, to symbolic (midi) files.
For the polyphonic transcription to midi files,
we used
the Onsets and Frames transcription system
\cite{onsets:frames:arxiv:2018},
%\cite{onsets:frames:blog},
developed by
the Magenta project.
%\cite{onsets:frames:blog},
It uses both convolutional
    and recurrent (LSTM) neural networks in order to: 1) predict pitch onset events; and
    %to
    2) use
%   these predictions
    this knowledge to condition framewise pitch predictions.
Obviously, we may also use instead directly MIDI music scores from existing symbolic music libraries.

\subsection{Layering}

We consider layers of music,
%selected
%three basic musical layers,
analogous to the production of orchestral music for games \cite{stuart:video:game:music:theguardian:2019},
with currently up to 4 layers:

\begin{itemize}

\item 1st layer, the most conservative and neutral;

\item 2nd layer, to add more excitement, e.g., though some additional instrument;

\item 3rd and 4th layers, to intensify the immersion and the tension.

\end{itemize}

\begin{figure}[htbp]
\begin{center}
\includegraphics[width=15cm]{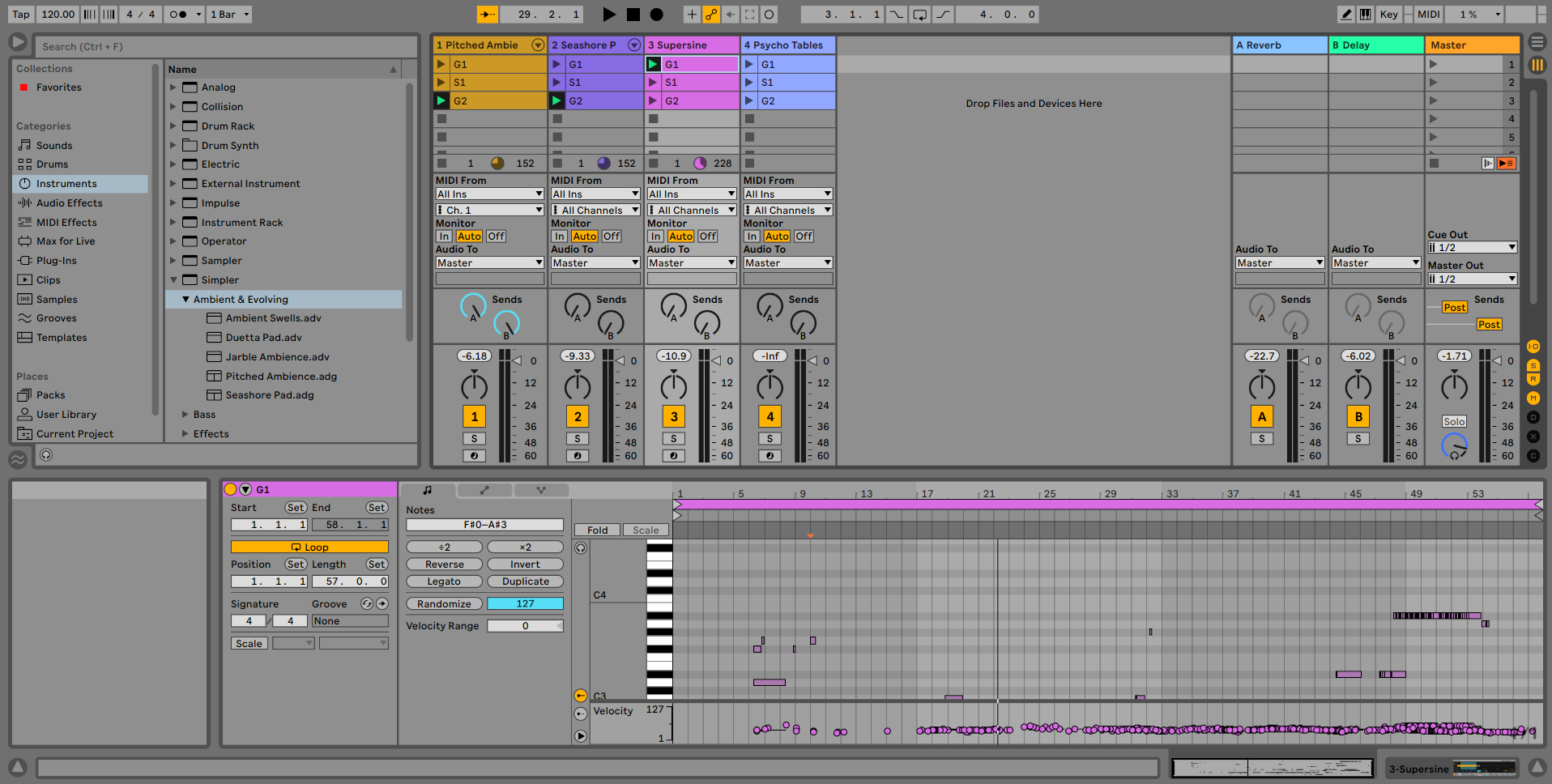}
\caption{The 4 Layers (each one with a different color) in the Ableton Live session view window}
\label{figure:layers:ableton}
\end{center}
\end{figure}

These layers are generated from the same learning corpus,
but from different seeds (starting sequences) and with different generation parameters
(currently,
    we vary a temperature parameter that controls the determinism of the generation, for some more likely or more unpredictable result),
depending on the controlling model (as will be presented in Section~\ref{section:complete:model}).
In addition to this static parameterization of their generation according to the controlling model,
each musical layer will be dynamically activated and played (or not)
(currently within the Ableton Live platform,
    a real-time sequencer for live music creation and production
    \cite{ableton:live},
see
%them in the Ableton Live sequencer in
Fig.~\ref{figure:layers:ableton}),
depending on the strategies of the controlling model.
%, as will be detailed in Section~\ref{section:strategy}.
%analogous to the production of orchestral music for games \cite{stuart:video:game:music:theguardian:2019}.

\subsection{Mapping Emotions}
\label{section:mapping:emotions}

In order to have some high-level and human understandable control of the generation by the game play context (game and player(s)),
we chose an emotion model, more precisely the arousal/valence model \cite{russell:circumplex:model:1980},
in which an emotion can be mapped using two parameters:

\begin{itemize}

\item the {\em arousal}, which represents the intensity of the emotion;
    
\item and the {\em valence}, which represents its quality (e.g., if it is positive, negative, neutral\ldots).

\end{itemize}

\begin{figure}[htbp]
\begin{center}
\includegraphics[width=10cm]{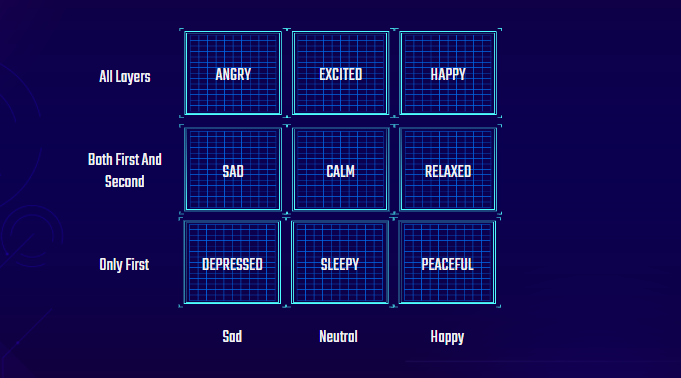}
\caption{Strategy/Layer/Emotion model,
with the 9 pre-defined emotions based on the arousal/valence emotion model}
\label{figure:layers:arousal:valence:model}
\end{center}
\end{figure}

In order to simplify our current prototype,
we now consider only 9 (discrete) emotions, as illustrated in
%Fig.~\ref{figure:selected:arousal:valence:model}.
Fig.~\ref{figure:layers:arousal:valence:model}.

%\begin{figure}[htbp]
%\begin{center}
%\includegraphics[width=4.5cm]{selected-arousal-valence-model.png}
%\caption{Arousal/valence emotion model and the 9 pre-defined emotions}
%\label{figure:selected:arousal:valence:model}
%\end{center}
%\end{figure}

The emotion model is designed as a server receiving control information from the game,
in order to be able to work with various games and game values models.
The game play information (events) emitted by the game may be about the game situation, player(s) situation,
but also from various other sources such as quests, terrains, etc.
How to aggregate these various informations is still an open issue for future work (see Section~\ref{section:game:music:mapping:model}).

\subsection{Strategy and Control Model}
\label{section:strategy}
\label{section:complete:model}

\begin{figure}[htbp]
\begin{center}
\includegraphics[width=15cm]{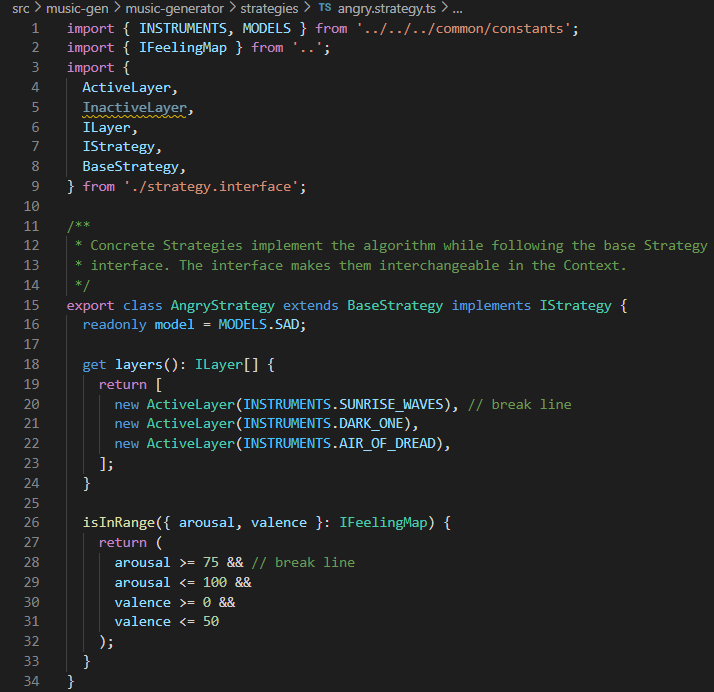}
\caption{Adding a new strategy named angry.
    It specifies:
    the instruments used for each layer (method {\tt get});
    and the range of arousal/valence values for triggering the strategy
    (method {\tt isInRange})}
\label{figure:strategy:code}
\end{center}
\end{figure}

While planning for the future some more advanced state machine for mapping the emotions into generation control strategies
(as will be detailed in Section~\ref{section:game:music:mapping:model}),
in current prototype we have implemented 9 pre-defined strategies
(corresponding to the 9 emotions shown in
%Fig.~\ref{figure:selected:arousal:valence:model}),
Fig.~\ref{figure:layers:arousal:valence:model}
%: happy, neutral and sad,
with for each one different values corresponding to the parameters for the generation:
which layers are activated, which instruments
(sampled or synthetic sounds,
currently selected from
    some instruments library for ambient music
    within Ableton Live)
are used and which effects are used.
More strategies/types may be added by adding strategy classes to the implementation
(as shown in Fig.~\ref{figure:strategy:code}).

%\subsection{Complete Control Model}
%\label{section:complete:model}

The complete model (Strategy/Layer/Emotion) for controlling music generation is shown in Fig.~\ref{figure:layers:arousal:valence:model}.
Current mapping is as follows:
the strength (arousal) corresponds to the number of active layers,
while the quality (valence) corresponds to the choice of emotional modes of the generated musical components.

%\begin{figure}[htbp]
%\begin{center}
%\includegraphics[width=6cm]{layers-arousal-valence-model.png}
%\caption{Complete Strategy/Layer/Emotion model}
%\label{figure:layers:arousal:valence:model}
%\end{center}
%\end{figure}

\subsection{Architecture}
\label{section:architecture}

\begin{figure}[htbp]
\begin{center}
\includegraphics[width=12cm]{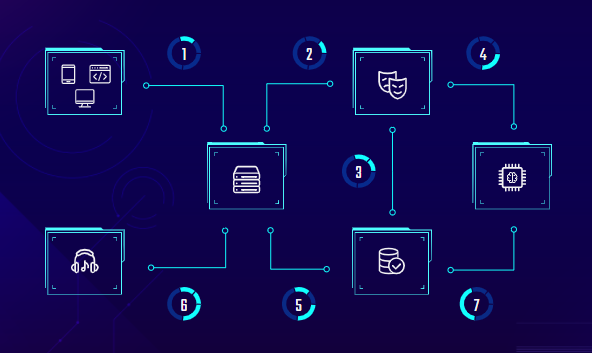}
\caption{Final architecture flow}
\label{figure:architecture:flow:2}
\end{center}
\end{figure}

%The architecture of current system is illustrated in Fig.~\ref{figure:architecture:flow:2}, and the flow logic is as follows:
The flow logic  of current architecture,
illustrated in Fig.~\ref{figure:architecture:flow:2},
is as follows:

\begin{enumerate}

\item User's client requests a music;

\item The server maps the user feeling through the arousal valence parameters;

\item It fetches, from memory, a song correspondent to the mapped emotion
(this optimization
    is
    %to be
    detailed in next Section~\ref{section:implemented:scheme});

\item If no associated music has already being generated, it starts the generation;
%of the most used layers;

\item After the music is fetched, it attaches metadata such as instruments;

\item It delivers the request response with the music to the final user;

\item The memory is refreshed.

\end{enumerate}

\subsection{Implementation}
\label{section:implemented:scheme}

To optimize the music generation process, at least one music corresponding to each strategy is saved in memory.
The architecture is designed as a server responsible for music generation,
for various possible game clients, based on game engines like Unity or Unreal,
or specific ones.
In order to automate and scale-up the machine learning life cycle,
we have used the Pachyderm platform (pipeline)
\cite{pachyderm}.
For the implementation, we have used Nvidia CUDA
development environment for high performance GPU-accelerated applications.
%\cite{whitaker:ambient:generator:2020}.

\subsection{Evaluation}
\label{section:evaluation}

Current architecture has been tested with an emulated game model and with music generated from a corpus of ambient music.
The complete code
	as well as input and output examples are available on the following code repository: {\tt https://gitlab.com/music-gen/server}.
Arousal valence values have been estimated according to possible moments of the hero's journey and the behavior of the system.
We are planning the integration with a real game using Unity.

\section{Prospects}
\label{section:prospects}

\subsection{Game/Music Mapping Model}
\label{section:game:music:mapping:model}

At present time, input from the game play is limited, but it could benefit from many more parameters and events 
(e.g., plot situation, player situation, including statistics, e.g., average player reactivity\ldots)
and how to aggregate them.
And, as opposed to mapping music to player state, we may want to oppose it instead, e.g., if the player is perceived to be showing to signs of abandon, you may want him to try to boost him with some positive music.
Last, note that \cite{plut:music:matters:cog:2019} proposes an additional dimension:
tension, that you could compute and use to improve the system's emotion mapping.

%\subsection{Use Of Image Recognition}
%
%Another really nice option for optimizing the feeling map could be the use
%of image recognition to predict players' emotions, or at least assist in the mapping process.

%\subsection{Multiple Input Sources}

%\subsection{Multiple Levels State Machine}

In addition,
as mentioned in Section~\ref{section:strategy},
we are planning to substitute current strategy scheme with some more abstract and general state machine model,
analog to the AMS spreading activation model (see Section~\ref{section:architecture:ams} and \cite{hutchings:amc:trans:2020}),
in order to track the transitions of the player's emotions.
Better transitions between music could also be planned ahead, through interpolation.

%Even more, with the implemented state machine, you can implement
%multi-level states, adding new abstraction levels to the models, for example, you
%could have a macro level for the game theme, a medium level for the music
%genre and a micro level for a more granular analysis of the player's own feeling.

%\subsection{Scale Up The System}

%Another future work is to scale up the system (for a large number of simultaneous players).
%The idea is to try to build an architecture capable of generating music
%for thousands of players at the same time, for an MMORPG for example.
%In this context, the server would have to be deployed in a cloud
%environment, but an in-depth analysis of what technology should be used, as
%there are many infrastructure technologies such as dockerized options like AWS
%ECS, FaaS options like Google Cloud Functions etc. that could be used in this
%process.

\subsection{Interactive Coordination}

\begin{figure}[htbp]
\begin{center}
\includegraphics[width=15cm]{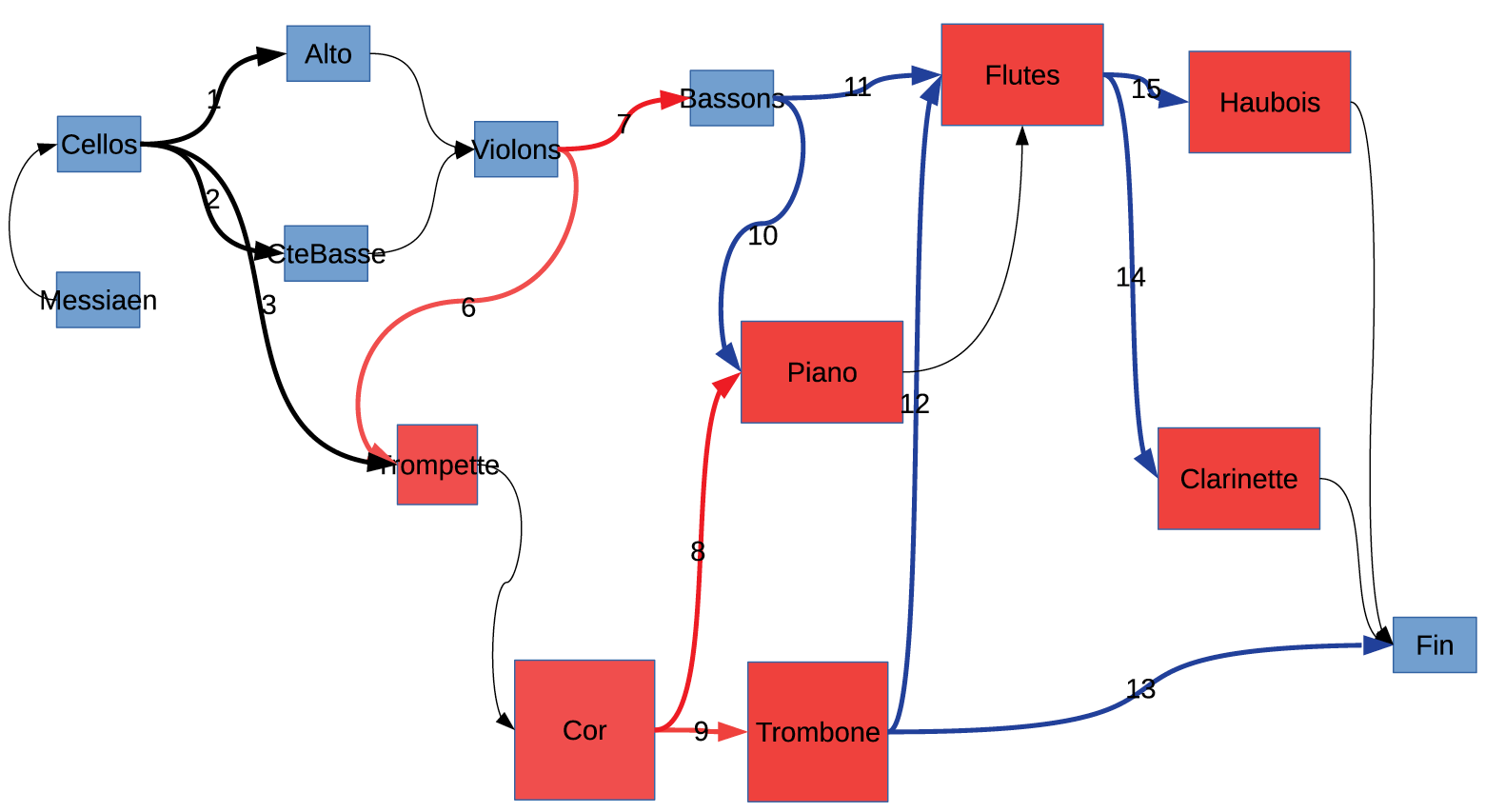}
\caption{Example of orchestral flow in Skini (Opus1 Piece by Bertrand Petit).
Plain arrows represent fixed paths and bold arrows represent alternative paths which may be decided by the public.
Each music/sound component (in blue) may be activated an unlimited number of times, except for ``reservoirs'' (in red) which are set to have some maximum number of activations}
\label{figure:opus1:automata}
\end{center}
\end{figure}

A more radical approach is to substitute the sequencer-like platform (currently, Ableton Live) by a more general platform for interactive and collaborative control of musical components (being generated by our current Transformer-based architecture).
We are thinking of the Skini platform
%of our colleague Bertrand Petit
\cite{petit:skini:asep:2020}.
%, with whom we have some collaboration.
%This platform
It allows defining some kind of ``orchestral blueprint'' (actually, some cartography of possible paths) for activating various musical components of a piece of music.
It separates the macro-level coordination from the actual micro-level components,
as for architectural/coordination languages in software architectures \cite{shaw:software:architecture:1996} or distributed systems.
Paths may be fixed
%(shown in plain arrows in Fig.~\ref{figure:opus1:automata})
or open with various choices,
%(shown in bold arrows),
to be decided according to the interaction with the public (various active listeners).
Fig.~\ref{figure:opus1:automata} shows an example of visual orchestral blueprint
(musical flow)
in Skini.

\begin{figure}[htbp]
\begin{center}
\includegraphics[width=10cm]{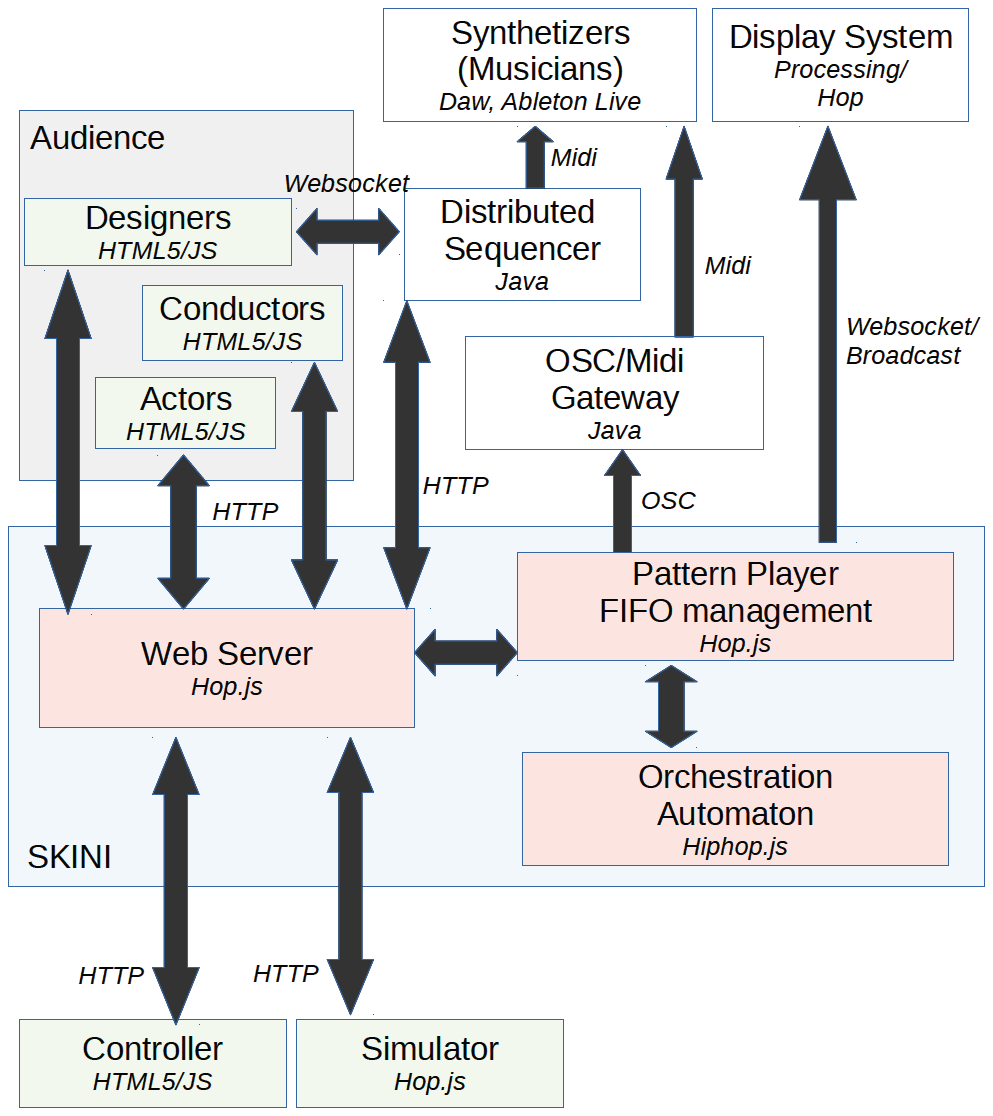}
\caption{Skini architecture}
\label{figure:skini:architecture}
\end{center}
\end{figure}

The control expression in the Skini platform is based on the integration of the synchronous reactive programming language Esterel \cite{berry:esterel:scp:1992} in JavaScript (on the Web).
The advantage over a sequencer (which has a semi-rigid temporal structure) is the expressive power (Turing complete) of a language like Esterel (which, for example, is used to control Airbus planes),
to program any type of coordination of real-time musical events, depending on various in-game events.
Additionally, Esterel has formal semantics and property verification
%proofing
tools, thus offering possibilities of formally verifying properties, such as the termination or non-simultaneous activation of two arbitrary musical components.
The Skini platform's capability for collaborative interactive orchestration (e.g., for simultaneous control interactions by several actors) offers us the right level of management of various interactions and controls
coming from the game engine and from the different players.
The Skini platform (whose architecture is shown in Fig.~\ref{figure:skini:architecture})
has already been tested recently, in a first scenario with a game platform (Unreal Engine 4),
to control the scheduling and musical orchestration depending on the situation of the player within the game \cite{petit:skini:game:2021}.

\section{Conclusion}
\label{section:conclusion}

In this paper, we have presented an architecture, based on deep learning (more specifically, the Transformer architecture),
for generating music for video games, personalized to the user musical preference.
It uses the technique of layering, with the activation of layers controlled by an emotion model,
in order to adapt it to the game play.
Our current architecture is a proof of concept, although it is complete and functional.
It has been tested with an emulated game model and with music generated from a corpus of ambient music.
We are currently working on the design of a next version architecture and its coupling with the coordination level based on the Skini architecture.
The objective is to decouple the generation and the adaptation of musical components from the way to coordinate and orchestrate them,
%according to the game play,
in order to refine the control of music adaptation according to the game play, independently of the music personalization.
We hope that the proposal, although preliminary, as well as the discussion and the prospects presented in this paper,
may humbly contribute to a better understanding of the issues and possible directions for next generation game music generation.

\subsubsection*{Acknowledgments}

We thank CNPq
(Conselho Nacional de Desenvolvimento Cient\'{\i}fico e Tecnol\'ogico, Brazil)
for
their financial support
through
Visiting Researcher (PV) fellowship/grant
N\textsuperscript{o} 302074/2020-1.

\bibliography{music-games}

\begin{thebibliography}{10}

\bibitem{ableton:live}
{Ableton team}.
\newblock Ableton {L}ive.
\newblock {A}bleton, Last access on 16/06/2022.
\newblock {M}usic creation and performance software.

\bibitem{berry:esterel:scp:1992}
G\'erard Berry and Georges Gonthier.
\newblock The {E}sterel synchronous programming language: Design, semantics,
  implementation.
\newblock {\em Science of Computer Programming}, 19(2):87--152, 1992.

\bibitem{starcraft:ii}
{Blizzard Team}.
\newblock Star{C}raft {II}: {W}ings of {L}iberty.
\newblock Blizzard Entertainment, 2010.
\newblock {G}ame.

\bibitem{nn4music:ncaa:2020}
Jean-Pierre Briot.
\newblock From artificial neural networks to deep learning for music generation
  -- {H}istory, concepts and trends.
\newblock {\em Neural Computing and Applications (NCAA)}, (33):39--65, January
  2021.
\newblock {S}pecial issue Neural networks in Art, sound and Design, J. Romero
  and P. Machado, Eds.

\bibitem{dlt4mg:springer:2019}
Jean-Pierre Briot, Ga{\"e}tan Hadjeres, and Fran{\c{c}}ois-David Pachet.
\newblock {\em Deep Learning Techniques for Music Generation}.
\newblock Computational Synthesis and Creative Systems. Springer, 2019.

\bibitem{collins:activation:1975}
Allan~M. Collins and Elizabeth~F. Loftus.
\newblock A spreading-activation theory of semantic processing.
\newblock {\em Psychological Review}, 82(6):407--428, 1975.

\bibitem{collins:interactive:audio:games:2011}
Karen Collins.
\newblock {\em From {P}ac-{M}an to Pop Music Interactive Audio in Games and New
  Media}.
\newblock Ashgate Publishing Ltd, 2011.

\bibitem{goodfellow:deep:learning:book:2016}
Ian Goodfellow, Yoshua Bengio, and Aaron Courville.
\newblock {\em Deep Learning}.
\newblock MIT Press, 2016.

\bibitem{hadjeres:anticipation:rnn:arxiv:2017}
Ga{\"e}tan Hadjeres and Frank Nielsen.
\newblock Interactive music generation with positional constraints using
  {A}nticipation-{RNN}, September 2017.
\newblock arXiv:1709.06404.

\bibitem{onsets:frames:arxiv:2018}
Curtis Hawthorne, Erich Elsen, Jialin Song, Adam Roberts, Ian Simon, Colin
  Raffel, Jesse Engel, Sageev Oore, and Douglas Eck.
\newblock Onsets and frames: Dual-objective piano transcription, June 2018.
\newblock arXiv:1710.11153.

\bibitem{lejaren:illiac:book:1959}
Lejaren~A. Hiller and Leonard~M. Isaacson.
\newblock {\em Experimental Music: Composition with an Electronic Computer}.
\newblock McGraw-Hill, 1959.

\bibitem{huang:music:transformer:arxiv:2018}
Cheng-Zhi~Anna Huang, Ashish Vaswani, Jakob Uszkoreit, Noam Shazeer, Ian
  Simon~Curtis Hawthorne, Andrew~M. Dai, Matthew~D. Hoffman, Monica Dinculescu,
  and Douglas Eck.
\newblock Music {T}ransformer: Generating music with long-term structure,
  December 2018.
\newblock arXiv:1809.04281.

\bibitem{jacopin:music:tempo:tetris:iui:2021}
Aline Hufschmitt, St\'{e}phane Cardon, and \'{E}ric Jacopin.
\newblock Dynamic manipulation of player performance with music tempo in
  {T}etris.
\newblock In {\em 26th International Conference on Intelligent User
  Interfaces}, IUI '21, pages 290--296, College Station, TX, USA, 2021. ACM.

\bibitem{hutchings:improvise:2017}
Patrick Hutchings and Jon McCormack.
\newblock Using autonomous agents to improvise music compositions in real-time.
\newblock In Jo{\~a}o Correia, Vic Ciesielski, and Antonios Liapis, editors,
  {\em Computational Intelligence in Music, Sound, Art and Design -- 6th
  International Conference, EvoMUSART 2017, Amsterdam, The Netherlands, April
  19--21, 2017, Proceedings}, volume 10198 of {\em LNCS}, pages 114--127.
  Springer, 2017.

\bibitem{hutchings:amc:trans:2020}
Patrick Hutchings and Jon McCormack.
\newblock Adaptive music composition for games.
\newblock {\em IEEE Transactions on Games}, 12(3):270--280, 2020.

\bibitem{layering:hyperbits:2022}
Hyperbits.
\newblock Layering music: 20 ways to layer sounds.
\newblock {H}yperbits, Last access on 16/06/2022.
\newblock {B}log.

\bibitem{jeffries:musician:machine:blog:magenta:2020}
Dan Jeffries.
\newblock The musician in the machine.
\newblock {M}agenta blog, August 2020.

\bibitem{jeffries:ambient:playlist:2022}
Dan Jeffries.
\newblock Ambient songs for creativity and calm.
\newblock Spotify, 2022.
\newblock {P}laylist.

\bibitem{juslin:music:emotions:2013}
Patrik~N. Juslin.
\newblock What does music express? basic emotions and beyond.
\newblock {\em Frontiers in Psychology}, 4:Article 596, 2013.

\bibitem{magenta}
Magenta.
\newblock Make music and art using machine learning.
\newblock Web Site.

\bibitem{nierhaus:algorithmic:composition:book:2009}
Gerhard Nierhaus.
\newblock {\em Algorithmic Composition: Paradigms of Automated Music
  Generation}.
\newblock Springer, 2009.

\bibitem{pachet:virtuoso:2012}
Fran{\c{c}}ois Pachet.
\newblock Musical virtuosity and creativity.
\newblock In Jon McCormack and Marc d'Inverno, editors, {\em Computers and
  Creativity}, pages 115--146. Springer, 2012.

\bibitem{pachet:markov:constraints:constraints:2011}
Fran{\c{c}}ois Pachet and Pierre Roy.
\newblock Markov constraints: Steerable generation of {M}arkov sequences.
\newblock {\em Constraints}, 16(2):148--172, 2011.

\bibitem{pachyderm}
{Pachyderm team}.
\newblock Pachyderm.
\newblock {G}ithub, Last access on 16/06/2022.
\newblock {C}ode documentation.

\bibitem{papadopoulos:maxorder:universality:book:2016}
Alexandre Papadopoulos, Fran{\c{c}}ois Pachet, and Pierre Roy.
\newblock Generating non-plagiaristic {M}arkov sequences with max order
  sampling.
\newblock In Mirko Degli~Esposti, Eduardo~G. Altmann, and Fran{\c{c}}ois
  Pachet, editors, {\em Creativity and Universality in Language}, Lecture Notes
  in Morphogenesis, pages 85--103. Springer, 2016.

\bibitem{papadopoulos:flow:composer:cp:2016}
Alexandre Papadopoulos, Pierre Roy, and Fran{\c{c}}ois Pachet.
\newblock Assisted lead sheet composition using {F}low{C}omposer.
\newblock In Michel Rueher, editor, {\em Principles and Practice of Constraint
  Programming: 22nd International Conference, CP 2016, Toulouse, France,
  September 5-9, 2016, Proceedings}, Programming and Software Engineering,
  pages 769--785. Springer, 2016.

\bibitem{petit:skini:game:2021}
Bertrand Petit.
\newblock {S}kini et jeu vid\'eo, January 2021.
\newblock {B}log.

\bibitem{petit:skini:asep:2020}
Bertrand Petit and Manuel Serrano.
\newblock {S}kini: Reactive programming for interactive structured music.
\newblock {\em The Art, Science, and Engineering of Programming}, 5(1):Article
  2, June 2020.

\bibitem{phi:transformer:towards:data:science:2020}
Michael Phi.
\newblock Illustrated guide to {T}ransformers -- step by step explanation.
\newblock {T}owards Data Science, April 2020.
\newblock {B}log.

\bibitem{plut:music:matters:cog:2019}
Cale Plut and Philippe Pasquier.
\newblock Music matters: An empirical study on the effects of adaptive music on
  experienced and perceived player affect.
\newblock In {\em 2019 IEEE Conference on Games (CoG)}, pages 1--8, 2019.

\bibitem{plut:music:games:ec:2020}
Cale Plut and Philippe Pasquier.
\newblock Generative music in video games: State of the art, challenges, and
  prospects.
\newblock {\em Entertainment Computing}, 33:100337, 2020.

\bibitem{ramalho:game:music:earphones:2021}
Geber Ramalho.
\newblock Communication during a debate/session.
\newblock Workshop on AI for (Music and Games) Co-Creation (WAIC 2021),
  November 2021.

\bibitem{russell:circumplex:model:1980}
James~A Russell.
\newblock A circumplex model of affect.
\newblock {\em Journal of Personality and Social Psychology}, 39(6):1161--1178,
  1980.

\bibitem{sanders:perception:music:video:games:bcs:2010}
Timothy Sanders and Paul Cairns.
\newblock Time perception, immersion and music in videogames.
\newblock In {\em Proceedings of the 24th BCS Interaction Specialist Group
  Conference}, BCS '10, pages 160--167, Swindon, U.K., 2010. BCS Learning \&
  Development Ltd.

\bibitem{shaw:software:architecture:1996}
Mary Shaw and David Garlan.
\newblock {\em Software architecture -- Perspectives on an emerging
  discipline}.
\newblock Prentice Hall, 1996.

\bibitem{zelda}
{Solarus Team}.
\newblock The {L}egend of {Z}elda: {M}ystery of {S}olarus.
\newblock Solarus, 2011.
\newblock {G}ame.

\bibitem{stuart:video:game:music:theguardian:2019}
Keith Stuart.
\newblock '{M}ozart would have made video game music': composer {E}{\'{\i}}mear
  {N}oone on a winning art form.
\newblock {\em The Guardian}, October 2019.

\bibitem{oord:wavenet:arxiv:2016}
A{\"a}ron van~den Oord, Sander Dieleman, Heiga Zen, Karen Simonyan, Oriol
  Vinyals, Alex Graves, Nal Kalchbrenner, Andrew Senior, and Koray Kavukcuoglu.
\newblock Wave{N}et: A generative model for raw audio, December 2016.
\newblock arXiv:1609.03499.

\bibitem{vaswani:attention:transformer:arxiv:2017}
Ashish Vaswani, Noam Shazeer, Niki Parmar, Jakob Uszkoreit, Llion Jones,
  Aidan~N. Gomez, Lukasz Kaiser, and Illia Polosukhin.
\newblock Attention is all you need, December 2017.
\newblock arXiv:1706.03762.

\bibitem{luftrausers:2014}
Vlambeer.
\newblock Luftrausers.
\newblock Devolver Digital, 2014.
\newblock {G}ame.

\bibitem{weir:no:man:sky:2017}
Paul Weir.
\newblock The sound of 'no man's sky', 2017.
\newblock {T}alk.

\bibitem{wilson:xcs:1995}
Stewart~W. Wilson.
\newblock Classifier fitness based on accuracy.
\newblock {\em Evolutionary Computation}, 3(2):149--175, 1995.

\end{thebibliography}
\bibliographystyle{plain}

\end{document}